\documentclass[aps,pra,twocolumn,superscriptaddress]{revtex4-1}

\usepackage{graphicx}
\usepackage{dcolumn}
\usepackage{bm}
\usepackage{amsmath}
\usepackage{amssymb}
\usepackage{latexsym}
\usepackage{epsfig}
\usepackage{amsbsy}
\usepackage{array}
\usepackage{amssymb}
\usepackage{setspace}
\usepackage{bm}
\usepackage{epstopdf}
\usepackage{subfigure}
\usepackage{color}

\def\sint{\ifmmode{- \!\!\!\!\!\! \int}
    \else{\hbox{$- \!\!\!\! \int \ $}}\fi}

\begin{document}

\title{Quantum Zeno effect in self-sustaining systems: suppressing phase diffusion via repeated measurements}

\author{Wenlin Li}
\affiliation{Physics Division, School of Science and Technology, University of Camerino, I-62032 Camerino (MC), Italy}
\author{Najmeh Es'haqi-Sani}
\affiliation{Physics Division, School of Science and Technology, University of Camerino, I-62032 Camerino (MC), Italy}
\author{Wen-Zhao Zhang}
\affiliation{Department of Physics, Ningbo University, Ningbo 315211, China}
\author{David Vitali}
\affiliation{Physics Division, School of Science and Technology, University of Camerino, I-62032 Camerino (MC), Italy}
\affiliation{INFN, Sezione di Perugia, via A. Pascoli, I-06123 Perugia, Italy}
\affiliation{CNR-INO, Largo Enrico Fermi 6, I-50125 Firenze, Italy}

\date{\today}
\begin{abstract}
We study the effect of frequent projective measurements on the dynamics of quantum self-sustaining systems, by considering the prototypical example of the quantum Van der Pol oscillator. Quantum fluctuations are responsible for phase diffusion which progressively blurs the semiclassical limit cycle dynamics and synchronization, either to an external driving, or between two coupled self-sustained oscillators. We show that by subjecting the system to repeated measurements of heterodyne type at an appropriate repetition frequency one can significantly suppress phase diffusion without spoiling the semiclassical dynamics. This quantum Zeno-like effect may be effective either in the case of one or two coupled van der Pol oscillators, and we discuss its possible implementation in the case of trapped ions.
\end{abstract}
\maketitle
\section{Introduction}  
The influence of measurement on quantum systems is a fundamental feature of quantum mechanics with many nontrivial manifestations~\cite{Nielsen2000,Weedbrook2012,Aharonov1988}. One striking and well-known example is the so-called quantum Zeno effect, which predicts that frequent measurements can freeze quantum dynamics \cite{Misra1977}. In recent years, several in-depth explorations have been performed both theoretically and experimentally \cite{Facchi2001,Facchi2001PRL,Facchi2004,Hotta2004,Thapliyal2016,Dhar2006,Streed2006,Xu2011,Harrington2017,Naikoo2019,Zheng2013,Wuster2017,Froml2019,Alvarez2010} and different mechanisms related to the Zeno effect have been investigated such as the nonmonotonic dissipation process caused by a structured reservoir, the renormalization effect due to a strong system-detector interaction \cite{Hotta2004}, or its connection with the opposite phenomenon of the anti-Zeno effect \cite{Facchi2001,Facchi2001PRL,Thapliyal2016,Xu2011,Lassalle2018,Chaudhry2016,Zhang2018R}. In quantum information processing it has been related to decoherence-free subspaces~\cite{Facchi2002} and to decoherence control strategies~\cite{Hosten2006,Wang2008,Huang2008,Shao2010,Li2016}. The relationship between Zeno effect and system symmetry has been discussed in Refs.~\cite{Naikoo2019,Li2006}, while Ref.~\cite{Porras2013} has shown that the quantum Zeno effect can be realized even using classical resources.

Quantum self-sustaining systems whose classical steady state forms a limit cycle in phase space have recently attracted much attention in the field of quantum science~\cite{Roulet2018,Lee2013,Ludwig2013,Weiss2017}. A prerequisite for the occurrence of nontrivial physical phenomena such as quantum synchronization~\cite{Roulet2018,Kwek2018} and quantum time crystal~\cite{Richerme2017} is the appearance of stable, self-sustained oscillations and this generally requires the existence of nonlinearities in the system's dynamics. Suitable platforms for exploring these phenomena are nonlinear systems such as the van der Pol (vdP) oscillator~\cite{Lee2013,Witthaut2017,Navarrete-Benlloch2008,kato2019,Jessop2019} and optomechanical systems (OMSs)~\cite{Ludwig2013,Weiss2017,Mari2013,Heinrich2011,Ying2014,Li2017,Bemani2017,Li2020,Piergentili2020a,Piergentili2020b}, and interesting manifestations which have been investigated both theoretically and experimentally in this respect are mode competition among limit cycles~\cite{Kemiktarak2014}, multi-stability~\cite{Marquardt2006}, Hopf bifurcation~\cite{Ying2014,Piergentili2020b} and chaotic behavior~\cite{Bakemeier2015,lu2015}.

The exact solution of the dynamics of these quantum nonlinear systems is hard and in the literature various approximate treatments have been proposed: some mean-field treatments linearize the dynamics of the quantum fluctuations around the solution of the mean field classical nonlinear equations, so that the steady state of the system is a Gaussian state centered around the classical limit cycle~\cite{Ying2014,Li2017,Bemani2017,Wang2014,Meng2020,Huan2015,Qiao2020}. However, such kind of state cannot be maintained after a transient, and inevitably exhibits non-Gaussianity, as was revealed in previous works by means of simulations~\cite{Lee2013,Navarrete-Benlloch2008,kato2019,Ludwig2013,Weiss2016}. Recently, a new fruitful perspective has been introduced in Ref.~\cite{Navarrete-Benlloch2017}, which showed that the long-time stationary dynamics of the quantum self-sustaining systems is characterized by a phase diffusion process yielding a non-Gaussian steady state well described by an appropriate mixture of Gaussians, distributed over all the phases of the limit cycle.

Stimulated by this result and by the importance of controlling and reducing phase diffusion in self-oscillating systems, we study here the effect of many repeated measurements on the self-sustaining system, an aspect which has remained unexplored up to now.
More specifically, we investigate the effect of ideal heterodyne measurements and also of a dichotomic projective measurement which is implementable in the case of trapped-ion resonators, on the dynamics of single- and bi-partite quantum self-sustaining systems, by extending the approach of Navarrete-Benlloch \textit{et al}.~\cite{Navarrete-Benlloch2017} to the semiclassical regime. We consider a quantum vdP oscillator in the semiclassical regime \cite{Lee2013} and find that an appropriate measurement frequency will lead to a Zeno-like effect such that the phase diffusion of the system is suppressed. We then consider the case of two coupled vdP quantum oscillators, and show that repeated measurements can also optimize the quantum synchronization between the two vdP oscillators. We will also see that, in the limit of high-frequency measurements, the system state is never fully freezed and self-trapped, but randomly walks in phase-space.

The paper is organized as follows: In. Sec.~\ref{Measurement and phase dynamics of a van der Pol oscillator}, we investigate
the non-Gaussian dynamics and the effect of the repeated ideal heterodyne measurements on a quantum vdP oscillator in the semiclassical regime. In subsection~\ref{Quantum Zeno effect in self-sustained system}, we describe the measurement process and its interplay with the phase diffusion process ultimately related to the inherent time translation symmetry of the system. In Sec.~\ref{Dichotomic measurement}, we turn our attention to a dichotomic measurement and analyze its effect on phase diffusion. The study of the effect of the repeated measurements is then extended to the case of two coupled vdP resonators in Sec.~\ref{Measurement optimizing quantum synchronization}, where we see how one can enhance synchronization. We finally summarize the results in Sec.~\ref{conclusion}.

\section{Measurement and phase dynamics of a van der Pol oscillator}
\label{Measurement and phase dynamics of a van der Pol oscillator}
In order to investigate the effect of measurement on the phase dynamics of a self-sustained system, we consider a prototypical self-sustained oscillator, i.e., a quantum vdP oscillator which undergoes two-excitation nonlinear loss and it is affected by a linear gain process. Its dynamics is described by the following master equation \cite{Lee2013,Navarrete-Benlloch2017}
\begin{equation}
\begin{split}
\dot{\rho}=-i[H,\rho]+\kappa_1\mathcal{L}[\hat{a}^\dagger]\rho+\kappa_2\mathcal{L}[\hat{a}^2]\rho,
\end{split}
\label{eq:master equation}
\end{equation}
where $H=\omega_m \hat{a}^\dagger\hat{a}$ ($\hbar=1$) is the Hamiltonian of a free oscillator and $\mathcal{L}[\hat{o}]\rho=2\hat{o}\rho\hat{o}^\dagger-\hat{o}^\dagger\hat{o}\rho-\rho\hat{o}^\dagger\hat{o}$ is the standard form of the Lindblad superoperator. The second and the third terms on the right side of the above master equation describe the linear gain process and the two-excitation nonlinear dissipation process, respectively \cite{Lee2013}. In order to conveniently calculate and describe the dynamics of the vdP oscillator, here we adopt the Wigner representation by introducing the standard Wigner function \cite{book1}:
\begin{equation}
\begin{split}
W(\alpha,\alpha^{*},t)=\dfrac{1}{\pi^2}\int d^2z\chi_s(z,z^*,t)e^{-iz^*\alpha^*}e^{-iz\alpha},
\end{split}
\label{eq:r1}
\end{equation}
where $\chi_s(z,z^*,t)=\text{Tr}[\rho(t)e^{iz^*a^\dagger+iza}]$ is the symmetrically-ordered characteristic function. The quantum master equation~\eqref{eq:master equation} can be rephrased in terms of the following partial differential equation for the Wigner function~\cite{book1}
\begin{equation}
\begin{split}
\partial_t W=&\{(\partial_\alpha\alpha+\partial_{\alpha^*}\alpha^*)[i\omega_m-\kappa_1+2\kappa_2(\vert\alpha\vert^2-1)]\\&+\partial_\alpha\partial_{\alpha^*}[\kappa_1+2\kappa_2(2\vert \alpha\vert^2-1)]\\
&+\dfrac{\kappa_2}{2}(\partial^2_{\alpha}\partial_{\alpha^*}\alpha+\partial_{\alpha}\partial^2_{\alpha^*}\alpha^*)\}W,
\end{split}
\label{eq:mean-field equation}
\end{equation}
where the higher order diffusion terms ($\partial^2_{\alpha}\partial_{\alpha^*}$ and $\partial_{\alpha}\partial^2_{\alpha^*}$) provide the possibility of a negative Wigner function, which is a manifestation of non-classical properties of the oscillator state. Quantum fluctuations are amplified by the nonlinear dissipation term with rate $\kappa_2$, and for not too small values of the ratio $\kappa_2/\kappa_1$ the dynamics of the system may significantly deviate from the classical limit cycle dynamics. However, here we focus on the semiclassical regime characterized by $\kappa_2/\kappa_1 \ll 1$~\cite{Lee2013}: in this limit derivatives of order higher than the second one can be neglected~\cite{Lee2013,book1}, and if the initial Wigner function is non-negative, it will remain non-negative at all times. In this case, the exact quantum dynamics described by Eq.~\eqref{eq:mean-field equation} can be well described by the following stochastic Langevin equation~\cite{book1}
\begin{equation}
\begin{split}
\dot{\alpha}=(-i\omega_m+\kappa_1)\alpha-2\kappa_2 (\vert \alpha\vert^2-1)\alpha+\sqrt{3\kappa_1+2\kappa_2}\alpha^{in}.
\end{split}
\label{eq:stochastic Langevin equation}
\end{equation}
In this expression, the first term on the right-hand side of the equation describes the oscillation frequency and gain of the vdP oscillator. The second term corresponds to the nonlinear dissipation, and the last term describes a stochastic fluctuation process, where $\alpha^{in}(t)$ is the Gaussian vacuum input noise acting on the oscillator~\cite{Lee2013,Li2017,Weiss2016}, with correlation function
\begin{equation}
\begin{split}
\langle \alpha^{in*} (t)\alpha^{in}(t')\rangle=\langle \alpha^{in}(t') \alpha^{in*}(t)\rangle=\delta(t-t').
\end{split}
\label{eq:stochastic N}
\end{equation}
The two quadratures of the vdP resonator are $Q=\alpha+\alpha^{*}$ and $P=i(\alpha^{*}-\alpha)$. Under this definition, the radius of the classical limit cycle $r$
is obtained from the stationary solution of the averaged Eq.~\eqref{eq:stochastic Langevin equation} and is given by
\begin{equation}
\begin{split}
r=2\vert\alpha_s\vert=2\sqrt{\dfrac{\kappa_1}{2\kappa_2}+1},
\end{split}
\label{eq:radius vdp}
\end{equation}
where $\vert \alpha_s\vert^2=\kappa_1/2\kappa_2+1$ can be regarded as the steady-state number of phonons. In order to numerically reproduce the essential aspects of the linearization theory of Ref.~\cite{Navarrete-Benlloch2017} around a nontrivial steady state such as a limit cycle, the initial state should be taken as a Gaussian state centered not too far from the limit cycle in order to ensure that the quantum fluctuations can be analyzed in terms of the Floquet theory approach of Ref.~\cite{Navarrete-Benlloch2017} so that the system stability and phase diffusion are determined by the corresponding Floquet eigenvalues rather than the instantaneous ones.
Therefore we numerically solve Eq.~(\ref{eq:stochastic Langevin equation}) by selecting as initial state of the system a coherent state with amplitude $r/2$, $\vert r/2\rangle$. As a consequence, the dynamics of the vdP oscillator is obtained by simulating a large number of stochastic trajectories $\alpha^j$, each starting from the complex value $\alpha^j(0)=r/2+\mathcal{Z}$, where $\mathcal{Z}$ is a Gaussian random complex number satisfying the normal distribution $\mathcal{N}(0,0.5)$ \cite{exp2}.  The ensemble-averaged amplitude and its quantum fluctuation are calculated as $\langle a\rangle(t)=\sum_{j=1}^N a^j(t)/N$ and $\langle \delta a^2\rangle(t)=\sum_{j=1}^N (a^j(t)-\langle a\rangle(t))^2/N$, where the superscript $j$ denotes the $j$th trajectory of the simulation~\cite{Li2017}.  The Wigner function of the system can be analyzed numerically by discretizing the continuous phase space into a series of square grids, and the corresponding side length $h$ can be regarded as the the size of the pixel in phase space.  According to these simulation results, it is convenient to count $N_{Q,P}$ which is the number of results satisfying $Q^j\in(Q-h/2,Q+h/2]$ and $P^j\in(P-h/2,P+h/2]$ at a given time. As mentioned above, in the considered semiclassical regime, the Wigner function is always non-negative and it behaves as a standard probability distribution in phase-space, which is obtained as
\begin{equation}
\begin{split}
W(Q ,P)=\lim_{h\rightarrow 0}\dfrac{N_{Q,P}}{N h^2}.
\end{split}
\label{eq:phase distribution function}
\end{equation}

We now subject the vdP resonator to a series of repeated measurements: during a time interval $t$, $n$ measurements are performed so that the measurements are evenly spaced in time by an interval $\Delta t=t/n$. We first consider the case of ideal heterodyne measurements, corresponding to projections onto the overcomplete basis of coherent states $E(\alpha)=\pi^{-1/2}\vert\alpha\rangle \langle\alpha\vert$, $\alpha \in \mathbb{C}$~\cite{Weedbrook2012,Yuen1980,book1}. In the ideal case, the state is projected onto a generic pure coherent state $\vert \alpha_i\rangle$ with a probability which is given by $P_i=\text{Tr}[E(\alpha_i)\rho E^\dagger(\alpha_i)]$, and it corresponds to a strong measurement, in which the interaction with the detection apparatus is able to project the system onto a pure state. In a more realistic scenario, the oscillator is projected onto a mixed state with phase space variances larger than the vacuum one, due to detection inefficiencies and added noise. The projection is far from ideal also in the case of weak measurements, which however will not be considered here. In our stochastic simulation, the ideal heterodyne measurement at time $n'\Delta t $ is described by randomly selecting an element $\alpha^k$ from the set $\alpha^k(n'\Delta t)\in\{\alpha^j(n'\Delta t),\forall j\}$, and choosing as initial value for the next time step after the measurement $\alpha^k(n'\Delta t)+\mathcal{Z}$, where again $\mathcal{Z}$ is a Gaussian random complex number with normal distribution $\mathcal{N}(0,0.5)$.

\begin{figure}[]
\centering
\includegraphics[width=3.3in]{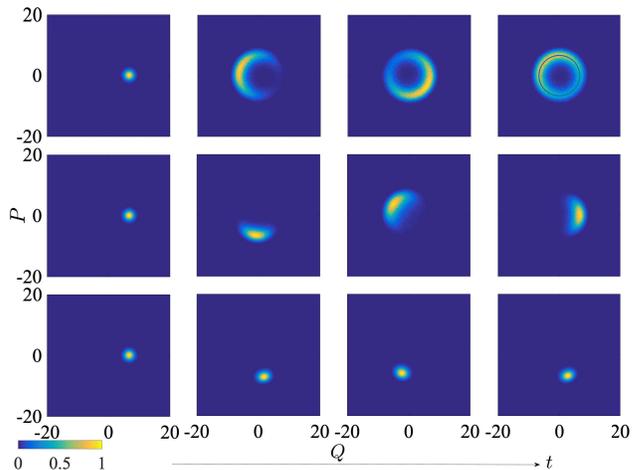}
\caption{Normalized Wigner function $W/\max\{W\}$ of the vdP oscillator in the presence of repeated ideal heterodyne measurements. The four columns from left to right correspond to $\omega_m t=0$, $\omega_m t=60$, $\omega_m t=120$ and $\omega_m t=180$, respectively. The first row represents the case without measurements ($\Delta t=\infty$), while the second and the third rows represent the dynamics in the presence of heterodyne measurement separated by a time interval $\omega_m\Delta t=10$ and $\omega_m\Delta t=1$ respectively. Each Wigner function is obtained from $500000$ trajectory solutions of the stochastic Langevin equation~\eqref{eq:stochastic Langevin equation}. The other parameters are $\kappa_1/\omega_m=0.1$ and $\kappa_2/\omega_m=0.005$. The solid line in the last sub-figure from the first row is the corresponding classical limit cycle.
\label{fig:2}}
\end{figure}

A visual and effective description of the dynamics of system's state, is provided in Fig.~\ref{fig:2}, where the time evolution of the Wigner function of the vdP oscillator is plotted at three different heterodyne measurement rates. In the first row of Fig.~\ref{fig:2}, we show the time evolution without any measurement, in which the initial Gaussian state gradually diffuses along the classical trajectory of the limit cycle with increasing phase fluctuation, finally becoming a ring, as expected. In the second row the evolution in the presence of heterodyne measurements separated by a time interval $\omega_m\Delta t=10$ is considered. The phase diffusion starts to be suppressed and the range of nonzero values of the Wigner function winding the classical trajectory is compressed. In the third row, the measurement rate is increased by a factor 10 ($\omega_m\Delta t=1$), and in this latter case, phase diffusion is completely suppressed and the Wigner function shows almost no deviation from a Gaussian. Therefore, the Zeno-like effect of repeated ideal heterodyne measurements proves to be efficient in suppressing phase diffusion due to the quantum fluctuations.

\begin{figure}[]
\centering
\includegraphics[width=3.3in]{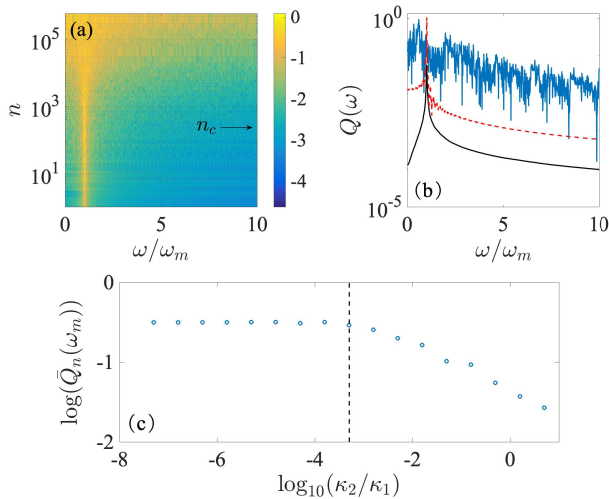}
\caption{(a) Fourier transform of the average quadrature $Q(\omega)$ as a function of the number of repeated heterodyne measurements $n$. (b) $Q(\omega)$ at three horizontal cuts of Fig.~\ref{fig:3}(a), corresponding to no measurement $n=0$ ($\Delta t = \infty$, black solid line), $n=16$ measurements ($\omega_m\Delta t=37.5$, red dash line), and $n=6\times 10^5$ heterodyne measurements ($\omega_m\Delta t=0.001$, blue ( upper) solid line). Here the ensemble average is obtained by $5000$ calculations of the stochastic Langevin equation and the other parameters are the same as those in Fig.~\ref{fig:2}. The arrow in (a) marks $n_c\sim 211$ which is the critical measurement rate corresponding to the threshold condition of
Eq.~(\protect\ref{eq:nc}). (c) Peak value of the Fourier transform, normalized with respect to the corresponding limit cycle radius, $Q_n(\omega_m)=Q(\omega_m)/r$, versus the rate ratio $\kappa_2/\kappa_1$, at fixed measurement rate $n=2000$, corresponding to $\omega_m\Delta t=0.3$ for a total evolution time $t=600$. The Fourier transform is obtained by  averaging $10$ times the result obtained with a single trajectory in order to eliminate the influence of randomness.
\label{fig:3}}
\end{figure}

The phase diffusion causes the expectation value of the oscillator's observables to average time-dependent oscillations to become constant. This phenomenon suggests us to characterize the Zeno-like process by means of the spectral analysis of the mean value of the oscillator's coordinate, namely,
\begin{equation}
\begin{split}
Q(\omega)=\left\vert\dfrac{1}{\sqrt{2\pi}}\int dt \langle Q\rangle(t)e^{-i\omega t}\right\vert,
\end{split}
\label{eq:Fourier transform}
\end{equation}
and the Fourier transform $Q(\omega)$ is shown in Fig.~\ref{fig:3}(a) versus the number of performed heterodyne measurements $n = t/\Delta t$, with a fixed total evolution time $\omega_m t=600$. In Fig.~\ref{fig:3}(b), we show three horizontal cuts of the spectra in Fig.~\ref{fig:3}(a), corresponding to no measurement $n=0$ ($\Delta t = \infty$, black solid line), $n=16$ measurements ($\omega_m\Delta t=37.5$, red dash line), and $n=6\times 10^5$ heterodyne measurements ($\omega_m\Delta t=0.001$, blue (upper) solid line). We see that if the measurement rate is not large enough the oscillatory behavior remains, together with a broadening due to phase diffusion. When the heterodyne measurements become frequent enough, they destroy the oscillations and a wide chaotic-like Fourier transform appear (see, e.g. the blue (upper) line in Fig.~\ref{fig:3}(b)). In this limit of large enough measurement rate, randomness accumulates and the motion of the oscillator explores a much wider phase space region compared to the limit cycle trajectory: in the numerical case of Fig.~\ref{fig:3} the size of the phase-space explored area is about $5$ times the classical limit cycle radius.

The transition from the oscillatory behavior to the chaotic motion caused by frequent enough heterodyne measurements is abrupt and one can define an effective measurement rate threshold $n_c$. Below this threshold, $Q(\omega)$ shows a single peak which even increases with increasing measurement rate (compare red and black lines in Fig.~\ref{fig:3}(b)) due to inhibition of phase diffusion caused by the measurements. Above the threshold, the Fourier transform becomes much wider and noisy. Therefore below the threshold the average phase shift between two successive measurements $\omega_m\Delta t $  is still distinguishable and larger than the accumulated phase dispersion caused by phase diffusion and that can be estimated as $2 \pi/r$. Therefore, the threshold between the two regimes can be identified as when the limit cycle phase shift is larger than three times the phase standard deviation, i.e., $3/r>\omega_m\Delta t /2\pi$, yielding the threshold condition
\begin{equation}
\begin{split}
\Delta t_c=\dfrac{3\pi}{\omega_m}\sqrt{\dfrac{2\kappa_2}{\kappa_1+2\kappa_2}},
\end{split}
\label{eq:nc}
\end{equation}
corresponding to the critical measurement rate $n_c=t/\Delta t_c$, which for the parameters used in Fig.~\ref{fig:3} is $n_c \simeq 211$.

The presence of a phase-transition-like behavior is confirmed by Fig.~\ref{fig:3}(c), where we have considered a fixed measurement time interval $\omega_m\Delta t=0.3$ ($n = 2000$), and plotted the peak value of the Fourier transform, normalized with respect to the corresponding limit cycle radius, $Q_n(\omega_m)=Q(\omega_m)/r$, versus the rate ratio $\kappa_2/\kappa_1$. The data show two well distinct behaviors: at small $\kappa_2/\kappa_1$ the normalized peak is constant, while it shows a power law decay as soon as $\kappa_2/\kappa_1$ becomes larger than a value which exactly corresponds to the threshold condition of Eq.~\eqref{eq:nc}: in this regime, phase diffusion due to the quantum fluctuations becomes dominant and the oscillatory behavior is progressively lost.

\subsection{General features of the quantum Zeno effect in self-sustained systems}
\label{Quantum Zeno effect in self-sustained system}
The inhibition of phase diffusion in a quantum vdP oscillator by means of repeated heterodyne measurements shown in the previous section when the measurement rate is properly chosen can be seen as a manifestation of a quantum Zeno-like process. We now discuss in more detail if and when this phenomenon is generalizable to any self-sustained system, and also the connection with the Zeno effect in general.

\begin{figure}[]
\centering
\includegraphics[width=3.5in]{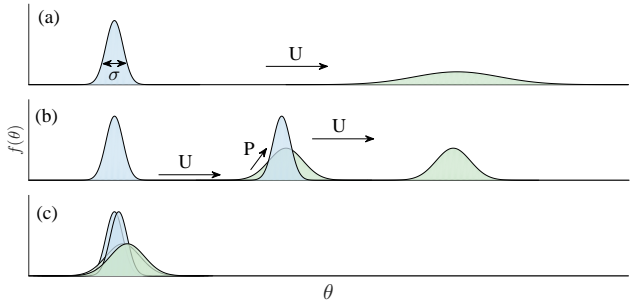}
\caption{Schematic representation of the interplay between phase diffusion and repeated measurements in quantum self-sustained system. (a): The dynamical process when no measurement is performed. (b): The case when measurements are performed with a not too large rate so that the limit cycle phase shift between two measurement is distinguishable and not overwhelmed by phase diffusion. (c): The case when measurements are performed with a large rate so that the limit cycle phase shift is too small. Here U (P) represents the evolution (measurement) process.
\label{fig:1}}
\end{figure}

In a generic self-sustained system, time translation symmetry is broken and a limit cycle is formed in phase-space. At the same time, as shown in Ref.~\cite{{Navarrete-Benlloch2017}}, quantum fluctuations are responsible for a phase diffusion process which spoils the classical limit cycle, and the quantum steady-state of the system can be expressed as a mixture of Gaussians~\cite{{Navarrete-Benlloch2017}}
\begin{equation}
\begin{split}
W=\int^{2\pi}_0f(\theta)W_G(\theta),
\end{split}
\label{eq:mixture of Gaussians}
\end{equation}
where $W_G$ are localized Wigner functions corresponding to the Gaussian state obtained by a linearized theory with coefficient matrix depending upon the classical trajectory. These Gaussian states will gradually diffuse and lose their definite phase, as described by Eq.~\eqref{eq:mixture of Gaussians}: the distribution $f(\theta)$ will become broader and broader until it becomes a uniform distribution, as shown in Fig.~\ref{fig:1}(a), restoring the time translation symmetry in this way at the end ~\cite{Richerme2017,Else2016}.

If we now insert repeated ideal projective measurements within the time evolution of the system, as in the quantum Zeno effect, these measurements will tend to freeze the system dynamics, and one can have two different scenarios, depending upon the value of the measurement frequency compared to the typical timescales of the nonlinear dynamics of the self-oscillating system.

As we have seen in the previous section, if during the time interval between two successive measurements $\Delta t$, the phase shift due to the oscillation $\Delta \theta$ is still distinguishable and larger than the increase of the standard deviation of the phase due to phase diffusion, $\sigma(\Delta t)<\Delta \theta$, the repeated measurements will inhibit phase dispersion and will keep the distribution $f(\theta)$ close to a $\delta$-function, as shown in Fig.~\ref{fig:1}(b). As a consequence, the system's state will remain Gaussian and will rotate around the classical limit cycle. In this case the Zeno effect is effective in suppressing the unwanted phase diffusion. If the measurement rate is large and $\Delta t \to 0$, the phase shift induced by the oscillation will become too small and the state will tend to be frozen, as in the ideal projective Zeno effect, and as shown in Fig.~\ref{fig:1}(c). These arguments can be applied to a generic self-oscillating system; however, in the model studied here and discussed in the previous section, we do not have perfect freezing as in the case of the standard Zeno effect but rather a random walk in phase space because we considered heterodyne measurements which are still projective, but in the \emph{overcomplete} basis of the coherent states. Due to the nonzero overlap between different basis states, the quantum state may change even in the limit $\Delta t\rightarrow 0$, and this is responsible for the residual random walk in phase space (e.g. the blue (upper) line in Fig.~\ref{fig:3}(b)).

For what concerns time translation symmetry and its eventual breaking, we notice that in general adding repeated measurements as in the Zeno effect makes the situation more involved. The numerical analysis carried out here does not allow us to draw definitive conclusions. However, in the regime below the threshold found in the previous section where phase diffusion is suppressed and the limit cycle is restored, the periodicity is still the classical one, longer than the measurement frequency and we have an effective amplification of the period as it occurs in Floquet time crystals~\cite{Else2016}. On the other hand, when the measurement rate is above threshold, the random walk in phase space of the state seems to suggest that the system is not able to reach its steady state, as also suggested in Ref.~\cite{lemini2018}.

\section{Dichotomic measurements}
\label{Dichotomic measurement}

As shown in Ref.~\cite{Lee2013}, the quantum vdP oscillator can be experimentally implemented with a trapped ion, by means of appropriate drivings resonant with the ion motional sidebands. However, even though trapped ions are an exceptional platform for quantum state and process engineering~\cite{Leibfried2003}, implementing ideal projective heterodyne measurements on them is highly nontrivial. What is instead quite easily achievable using the electron shelving technique~\cite{Leibfried2003} is the
dichotomic measurement corresponding to the positive operator valued measurement (POVM) with measurement operators $\{M_1=\vert\alpha\rangle\langle \alpha\vert,M_2=I-\vert\alpha\rangle\langle \alpha\vert\}$, that is, the projectors corresponding to the results to the yes/no question if the oscillator is in the desired coherent state $\vert\alpha\rangle$ or not. It is therefore interesting to see if repeating measurements of this kind on the system is still effective in suppressing phase diffusion. Compared with the ideal projective heterodyne measurement, such a dichotomic measurement only determines whether a quantum state is on a selected coherent state $\vert\alpha\rangle$ or not. This POVM is obtained by applying a properly chosen phase space displacement operation to the POVM corresponding to establishing if the ion motional state is in its ground state $ \vert 0\rangle $ or not, and which is realized in trapped ions with the electron shelving technique~\cite{Leibfried2003}. Also this technique corresponds to a strong measurement, capable of projecting (in case of success) onto a pure state. In our model, the selected target state is a periodic rotating coherent state $\rho=\vert \alpha\rangle= \vert \sqrt{\text{Tr}(\rho_s\hat{a}^\dagger\hat{a})} e^{-i\omega_m t}\rangle$ centered at the trajectory of the limit cycle. After applying the displacement transformation $\rho'=D^\dagger(\alpha)\rho D(\alpha)$, the dichotomic measurement can be described as a POVM measurement on $\rho'$ with POVM operators $\{M_1=\vert 0\rangle\langle 0\vert, M_2=I-\vert 0\rangle\langle 0\vert\}$. The quantum state after a measurement $M_i$ can be expressed as
\begin{equation}
\begin{split}
\rho_{mi}=D(\alpha)\left[\dfrac{M_i\rho'M^\dagger_i}{\text{Tr}(M_i\rho'M^\dagger_i)}\right]D^\dagger(\alpha),
\end{split}
\label{eq:measure state}
\end{equation}
and the corresponding probability is $P_i=\text{Tr}(M_i\rho'M^\dagger_i)$.

\begin{figure}[]
\centering
\includegraphics[width=3in]{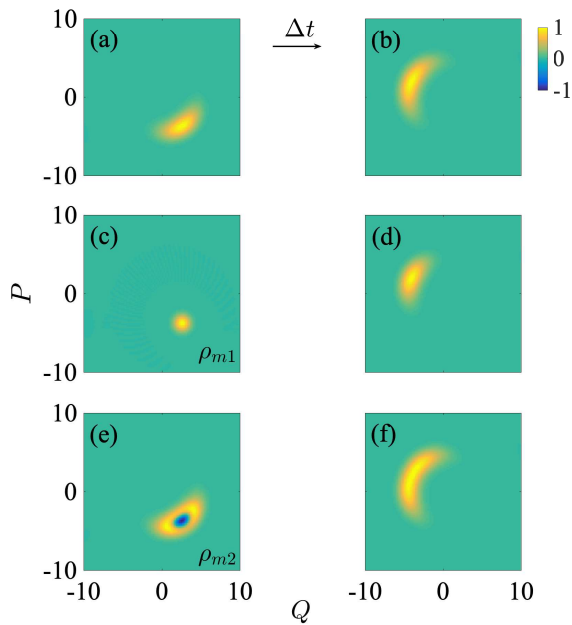}
\caption{Normalized Wigner function $W/\max\{W\}$ of vdP oscillator influenced by the dichotomic measurement. Evolution (a)$\rightarrow$(b): the phase of the system's state further diffuses during a standard time evolution of duration $\omega_m\Delta t=10$. Evolution (a)$\rightarrow$(c)$\rightarrow$(d): the POVM operator $M_1$ maps the phase diffused state to a coherent state $\rho_{m1}$ which further evolves to a state whose phase diffusion is suppressed. Evolution (a)$\rightarrow$(e)$\rightarrow$(f): the POVM operator $M_2$ maps the system's state to a ``complementary" state of the coherent state $\rho_{m2}$ with negative Wigner function, and phase diffusion is even wider after the evolution. The numerical calculation is performed in a $50 \times 50$ Hilbert subspace with Fock state basis and the other parameters are the same as those of Fig.~\ref{fig:2}.
\label{fig:7}}
\end{figure}

When a series of such dichotomic measurements are inserted into the evolution process of a vdP oscillator, the physical process corresponding to the POVM operator $M_1$ reproduces the desired action of the ideal projective measurement onto the overcomplete coherent state basis, that is, the measurement correctly projects the state onto a coherent state. In this case, as shown in Fig.~\ref{fig:7}(a), (c) and (d), the phase of this coherent state will diffuse again until a next successful measurement process $M_1$ eliminates the accumulated phase diffusion, and so on. However, in contrast to the complete heterodyne measurement of the previous section, and as shown in Fig.~\ref{fig:7}(e), the other POVM operator $M_2$ maps the system's state to a state with a distinct quantum nature, corresponding to a Wigner function with possibly negative values. It can be seen by comparing sub-figures (b), (d) and (e) that phase diffusion may be even magnified after that the oscillator is projected onto $\rho_{m2}$ instead of $\rho_{m1}$.

\begin{figure}[]
\centering
\includegraphics[width=3.3in]{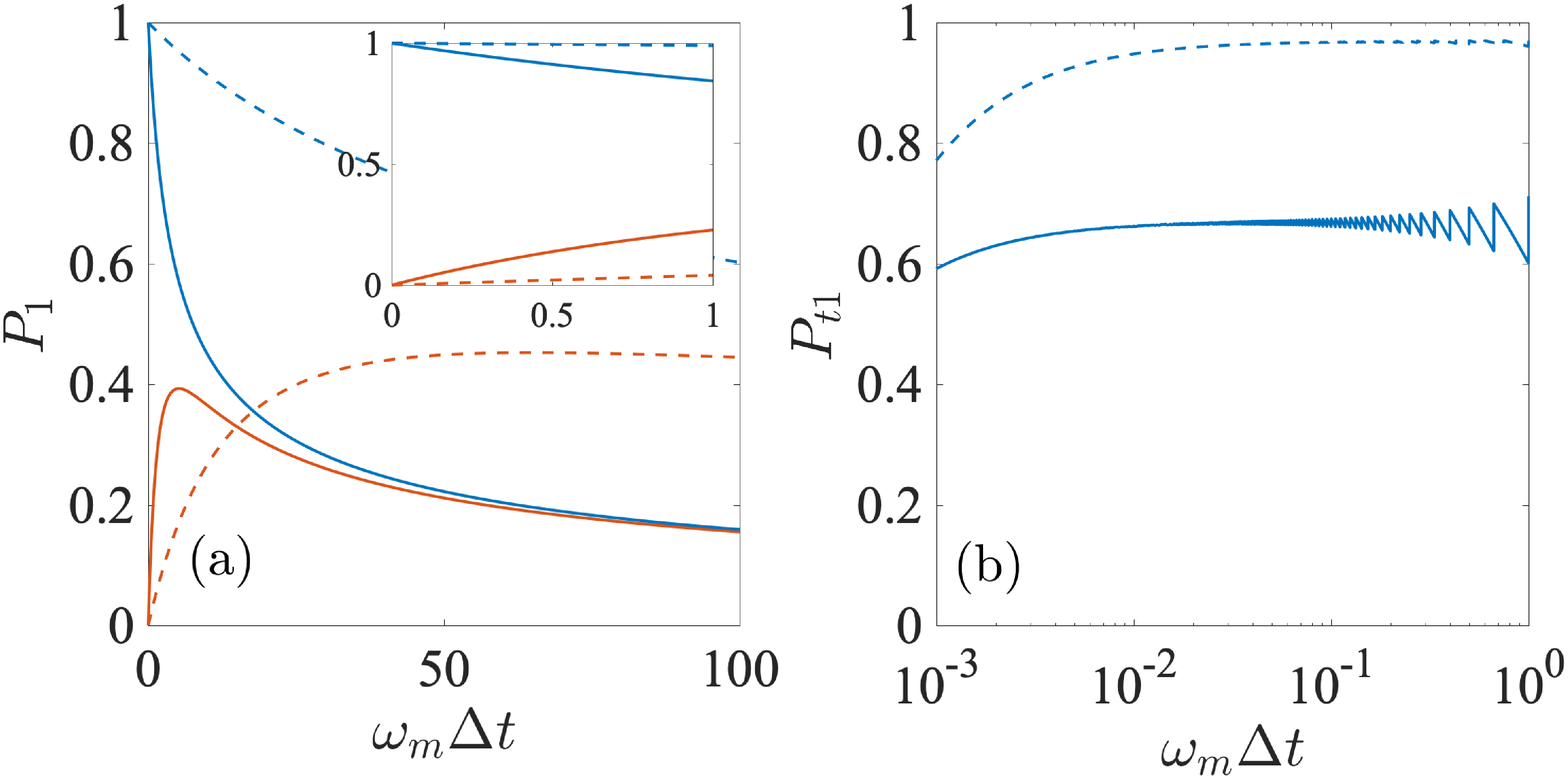}
\caption{(a) The probability $P_1$ of obtaining a coherent state in a dichotomic measurement as a function of the measurement time interval $\Delta t$. The blue (upper) line corresponds to the evolution of the system from a coherent state, i.e., the last measurement result is $\rho_{m1}$, while the red (lower) line refers to the case where the last measurement gave $\rho_{m2}$. (b) The probability $P_{t1}=(P_1)^n$ that the vdP oscillator stays in the desired coherent state in a given time interval $\omega_m t=2$ as a function of the time interval between two measurements $\Delta t$. In (a) and (b), the solid lines refer to $\kappa_1/\omega_m=0.1$ and $\kappa_2/\omega_m=0.005$ (the ''classical'' limit of a vdP oscillator), while dashed lines refer to $\kappa_1=0.005/\omega_m$ and $\kappa_2/\omega_m=0.01$.
\label{fig:8}}
\end{figure}

Despite the negative effect of the complementary unsuccessful measurement $M_2$, one can investigate if it is possible to find a regime where the probability to project onto the desired coherent state approaches one, so as to circumvent the unwanted effect of $M_2$. This analysis is carried out in Fig.~\ref{fig:8}. In Fig.~\ref{fig:8}(a) we plot the probability $P_1$ corresponding to the desired result in the $k+1$th dichotomic measurement versus the measurement interval $\Delta t$ under the condition that a coherent state is obtained at the generic $k$th measurement (full and dashed blue lines, referring to two different parameter regimes). It can be seen that the probability of obtaining again the coherent state will monotonically decrease with increasing $\Delta t$ due to the constant influence of phase diffusion by quantum noise, and therefore one needs small measurement intervals $\Delta t$ in order to try to suppress phase diffusion. For small $\Delta t$ $P_1$ linearly decreases, that is $P_1(\Delta t)=1-c\Delta t$ (see the inset of Fig.~\ref{fig:8}(a)). As a consequence, if one considers a given total time $t=n\Delta t$, the probability that the dichotomic measurements always give the same desired coherent state projection within this time interval is $P_{t1}=\left[P_1(\Delta t)\right]^n$,
\begin{equation}
\begin{split}
P_{t1}=\left[P_1(\Delta t)\right]^n\simeq \left(1-c \dfrac{t}{n}\right)^n\simeq e^{-c t}.
\end{split}
\label{eq:vdp probability}
\end{equation}
In this case, the factor $n$ is eliminated from the formula, which means that neither the Zeno effect nor the anti-Zeno effect occurs during the whole process, and therefore this dichotomic measurement is not able to suppress phase diffusion efficiently. We verify this assertion by examining the variation of $P_{t1}$ with the measurement frequency $n$ and plotting the results in Fig.~\ref{fig:8}(b). It can be seen that $P_{t1}$ will not get any significant improvement by increasing the measurement rate $n$ and therefore decreasing $\Delta t$ to low values.

We have also investigated if instead the oscillator can be frozen into some state, complementary to the coherent state, i.e., the state corresponding to the result of the POVM operator $M_2$. In this case however, the measurement projects onto a high-dimensional subspace and the resulting state after one $M_2$ measurement is generally different from the resulting state after the subsequent $M_2$ measurement. Therefore, there is no simple formula as that of Eq.~(\ref{eq:vdp probability}). However, through simulations we find that there is still no sign of the occurrence of the Zeno effect. This means that even though we repeat the evolution-measurement process, the state will not be frozen, but jumps between the coherent state and its corresponding ``complementary state"  take place. Nonetheless, compared with the steady-state $\rho_s$ with fully undetermined phase obtained without any measurement, we see that some inhibition of phase diffusion occurs also for the dichotomic measurement, although less efficient compared to the ideal projective heterodyne measurement.

\section{Measurement-enhanced quantum synchronization of two vdP oscillators}
\label{Measurement optimizing quantum synchronization}
In the previous sections, we showed how frequent ideal heterodyne measurements can suppress the phase diffusion on a self-oscillating system. It is interesting to see if in the case of a multipartite self-sustained system, these measurements can affect and eventually increase the phase correlations between the subsystems. It has been shown that two coupled quantum vdP oscillators can be spontaneously phase-correlated or synchronized~\cite{Lee2013,Lee2014}. In this work, the interaction between two oscillators is chosen to obey the $U(1)$ symmetry, so that the dynamics of the system can be described by the following master equation
\begin{equation}
\begin{split}
\dot{\rho}=-i[H,\rho]+\sum_{j=1,2}\kappa_1(\mathcal{L}[\hat{a}_j^\dagger]\rho)+\kappa_2\mathcal{L}[\hat{a}_j^2]\rho,
\end{split}
\label{eq:master equation two}
\end{equation}
with Hamiltonian $H=\sum_{j=1,2}\omega_{mj} \hat{a}_j^\dagger\hat{a}_j-\mu(\hat{a}^\dagger_1a_2+\hat{a}^\dagger_2a_1)$ with coupling strength $\mu$. The corresponding
stochastic Langevin equations, in the same semiclassical limit considered earlier, are given by
\begin{equation}
\begin{split}
\dot{\alpha}_j=&(-i\omega_ {mj}+\kappa_1)\alpha_j-2\kappa_2 (\vert \alpha_j\vert^2-1)\alpha_j\\
&+i\mu\alpha_{3-j}+\sqrt{3\kappa_1+2\kappa_2}\alpha_j^{in},~~~~~~j=1,2,
\end{split}
\label{eq:stochastic Langevin equations two}
\end{equation}
where the two input noises are uncorrelated Gaussian vacuum input noises of the identical to those introduced in Eq.~(\ref{eq:stochastic Langevin equation}), satisfying the correlation functions of Eq.~\eqref{eq:stochastic N}.

Let us first review the properties of the coupled vdP oscillator model for what concerns phase-correlations in the presence of quantum noise, which is usually quantified in terms of the probability distribution of the phase difference $\theta_-=\theta_1-\theta_2$, where we have defined $\alpha_j=I_je^{i\theta_j}$.  For this model, the $U(1)$ symmetry leads to a bistable distribution with equally probable in-phase and anti-phase synchronization~\cite{Lee2013}. This can be understood also analytically by first rewriting the stochastic equations in terms of the amplitude and phase variables,
\begin{equation}
\begin{split}
&\dot{I}_1=(\kappa_1-2\kappa_2)I_1-2\kappa_2 I_1^3+\mu I_2\sin\theta_-+N_{I_1},\\
&\dot{I}_2=(\kappa_1-2\kappa_2)I_2-2\kappa_2 I_2^3-\mu I_1\sin\theta_-+N_{I_2},\\
&\dot{\theta}_-=-\Delta \omega+\mu\left(\dfrac{I_2}{I_1}-\dfrac{I_1}{I_2}\right)\cos\theta_-+N_{\theta_-},
\end{split}
\label{eq:stochastic amplitude equations two}
\end{equation}
where $\Delta \omega=\omega_{m1}-\omega_{m2}$ is the frequency difference between the two oscillators. The terms $N_{I_{1,2}}=\sqrt{3\kappa_1+2\kappa_2}(\alpha_{1,2}^{in}e^{-i\theta_{1,2}}+\alpha_{1,2}^{in,*}e^{i\theta_{1,2}})$ are the amplitude noises and $N_{\theta_-}=\sqrt{3\kappa_1+2\kappa_2}[(\alpha_1^{in}e^{-i\theta_{1}}-\alpha_1^{in,*}e^{i\theta_{1}})/I_1-(\alpha_2^{in}e^{-i\theta_{2}}-\alpha_2^{in,*}e^{i\theta_{2}})/I_2]$ corresponds to the phase noise term responsible for phase diffusion. Then, in the weak coupling limit $\mu/\omega_{mj} \ll 1$, one can approximate the dynamics of two amplitudes $I_{1,2}$ as perturbations around their equilibrium positions at zero coupling, that is, $I_j=I_0+\delta I_j$, where $I_0=r/2=\sqrt{\kappa_1/2\kappa_2+1}$. We get
\begin{equation}
\begin{split}
&\delta \dot{I}_1=(\kappa_1-2\kappa_2)\delta I_1-6\kappa_2 I_0^2\delta I_1+\mu I_0\sin\theta_-+N_{I1},\\
&\delta \dot{I}_2=(\kappa_1-2\kappa_2)\delta I_2-6\kappa_2 I_0^2\delta I_2-\mu I_0\sin\theta_-+N_{I2},\\
&\dot{\theta}_-=-\Delta \omega+2\mu\left(\dfrac{\delta I_2-\delta I_1}{I_0}\right)\cos\theta_-+N_{\theta_-}.
\end{split}
\label{eq:delta stochastic amplitude equations two}
\end{equation}
Finally, one gets an effective equation for the phase difference $\theta_-$ only, by determining the stationary values of the amplitude fluctuations, obtained by setting the left-hand side of the first two equations in Eqs.~\eqref{eq:delta stochastic amplitude equations two} equal to zero. One obtains $\delta I_{3-j}=(-1)^j\mu I_0\sin\theta_-/(2\kappa_1+8\kappa_2)$, and using these expression into the third equation for $\theta_-$, one arrives at
\begin{equation}
\begin{split}
\dot{\theta}_-&=-\Delta \omega-\left(\dfrac{\mu^2}{2\kappa_1+8\kappa_2}\right)\sin2\theta_-+N_{\theta_-}\\
&\equiv -\Delta \omega-2c\sin2\theta_-+N_{\theta_-},
\end{split}
\label{eq:kuramoto}
\end{equation}
which is a Kuramoto-type equation with a washboard potential $U=-\int \dot{\theta}d\theta=\Delta \omega\theta_1-c\cos2\theta_-$ \cite{Marquardt2006,Weiss2016,explain1}. The $\sin 2\theta_-$ term is responsible for the presence of two local potential minima located at $\theta_-=0$ and $\pi$, respectively, clearly showing the bistability of the synchronized states, and the equiprobable in-phase and anti-phase correlated states.

\begin{figure}[]
\centering
\includegraphics[width=3.5in]{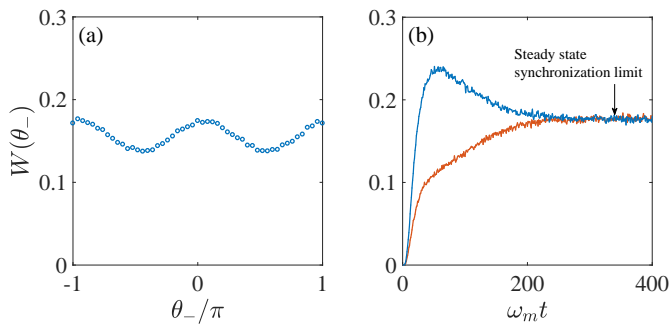}
\caption{(a): Phase error distribution of steady state for two coupled oscillators with $\omega_{m1}=1$, $\Delta \omega=0.01$ and $\mu=0.02$. (b): Time evolution of phase error distributions corresponding to in-phase synchronization (blue, upper) and anti-phase synchronization (red, lower). The results are obtained by $10^6$ calculations of the stochastic Langevin equations and other used parameters are the same with those in Fig.~\ref{fig:2}.
\label{fig:5}}
\end{figure}

We have numerically solved the stochastic Langevin equations~\eqref{eq:stochastic Langevin equations two} with the initial phase difference $\theta_-=\theta_1-\theta_2=\pi/2$ and we show the results in Fig. \ref{fig:5}. Fig.~\ref{fig:5}(a) shows the stationary probability distribution of the phase difference which can be obtained from the joint Wigner function expressed in polar coordinates ($I_1$,$I_2$,$\theta_+$,$\theta_-$) and integrating over the variables $I_1$, $I_2$ and $\theta_+$. We clearly see the two equal maxima at $\theta_-=0,\pi$. In Fig.~\ref{fig:5}(b) we plot the time evolution of the probability density at the two maxima, $W_{\theta_-}(0)$ and $W_{\theta_-}(\pi)$. The two oscillators tend firstly to in-phase synchronization, and then the degree of anti-phase synchronization gradually increases, resulting in a decrease in $W_{\theta_-}(0)$ until they reach the same steady-state value. This identical stationary value of the two maxima provides a quantification of the phase correlation (or anti-correlation) of the two oscillators and it is now interesting to study what is the effect of repeated ideal heterodyne measurements on this correlation.

\begin{figure}[]
\centering
\includegraphics[width=3in]{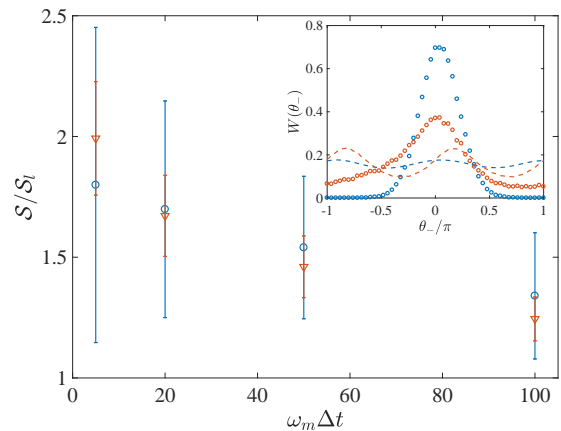}
\caption{Synchronization measure of Eq.~(\ref{eq:measure}) normalized by the value corresponding to the stationary values of Fig.~\ref{fig:5}(b), versus the measurement time interval $\Delta t$. Here, blue circles refer to the parameter choice $\Delta \omega/\omega_{m1}=0.01$ and $\mu/\omega_{m1}=0.02$, while the red triangles to the choice $\Delta \omega/\omega_{m1}=0.1$ and $\mu/\omega_{m1}=0.1$. Each $\mathcal{S}^i$ is obtained from $50000$ calculations of the stochastic Langevin equation and we average over $M=20$ values of $\mathcal{S}^i$ in order to get $\mathcal{S}$. The inset shows the phase difference distribution corresponding to the steady state without measurements for the two parameter choices (dashed lines), and to the case in the presence of repeated heterodyne measurements with a time interval $\omega_{m1}\Delta t=10$ (circles). The other parameters are the same as those of Fig.~\ref{fig:2}.
\label{fig:6}}
\end{figure}

We quantify the effect of the repeated measurements on the phase correlations of the two coupled oscillators by considering the time evolution of the probability density of the phase difference $\theta_-$, evaluated at the perfect correlation condition, $W_\theta(0)_t$, and at the anti-correlated condition, $W_\theta(\pi)_t$. We define the phase correlation at moment $t$ as the largest between the two values, and from this we define the degree of synchronization of the two coupled oscillators as the following time average, providing a robust signature of the phase correlation between the two coupled oscillators,
\begin{equation}
\begin{split}
\mathcal{S}^i=\dfrac{1}{T}\int^T_0dt \max[W^i_{\theta_-}(0)_t,W^i_{\theta_-}(\pi)_t].
\end{split}
\label{eq:measure}
\end{equation}
The repeated measurements bring extra randomness, and a steady state is not reached anymore in general because of the measurements, and this is why we average over time. Then we average also over $M=20$ repetitions and consider $\mathcal{S}=\sum \mathcal{S}^{i}/M$. We plot the numerical results for this quantifier in Fig.~\ref{fig:6}, where we have normalized $\mathcal{S}$ with respect to the steady-state value $\mathcal{S}_l=W(0) =W(\pi)$ of the unmeasured, symmetric case shown of Fig.~\ref{fig:5}(b). The associated standard deviation of this quantifier over the $M$ repetitions is $\sigma(\mathcal{S}/\mathcal{S}_l) \sim 0.3.$.

Fig.~\ref{fig:6} shows $\mathcal{S}/\mathcal{S}_l$ versus the measurement time interval $\Delta t$ for two different choices of the frequency difference $\Delta \omega$ and coupling rate $\mu$ (red plus and blue circles), and we see that quantum synchronization between the two vdP oscillators is enhanced over a wide range of the measurement time interval $\Delta t$. In particular, the largest improvement of phase correlation is obtained at short $\Delta t$, and this is somehow consistent with the time evolution of the phase difference distribution maxima in Fig.~\ref{fig:5}(b), where one has larger values for $W(0)$ at shorter times. This is also visible in the inset, where we compare the stationary phase difference probability distribution without measurements (dashed lines in the inset) with the one corresponding to the time averaged nonstationary function obtained with repeated ideal heterodyne measurements with $\omega_{m1}\Delta t = 10$ (the optimal case shown in the main figure), and corresponding to circles. In fact now only one kind of phase-correlation is enhanced by the measurement, and one has only one well distinct peak in the phase difference distribution, characterized by a smaller variance, and therefore an appreciably reduced phase diffusion.

\section{conclusion}
\label{conclusion}
In summary, we have investigated the influence of repeated measurements on the dynamics of quantum self-sustained systems. We have shown that the phase diffusion of the systems under study can be suppressed similarly to what occurs in the Zeno effect. In this case, by appropriately choosing the measurement rate, and in the case of ideal heterodyne projective measurements, the dynamics is stabilized around the classical limit cycle with a rotating Gaussian-like state with suppressed phase diffusion. Differently from the standard Zeno effect, if the measurement rate exceeds a critical value, the final system's state does not freeze at its initial state but is characterized by a random walk in phase space, due to the overcompletness of the coherent state basis.

Then we have studied the effect of repeated heterodyne measurements on the phase correlations between two coupled vdP oscillators. We found that these measurements improve the phase-correlations and enhance quantum synchronization with respect to the stationary value which is achieved in the model in the absence of any measurement.

We expect that the Zeno effect and the suppression of phase diffusion discussed in this paper are not restricted to vdP oscillator systems, but occur in any repeatedly measured self-sustaining system, provided that the kind of measurement and its rate are conveniently chosen. Nonetheless, in limit cycles time translation symmetry is broken and repeated measurements are an additional ingredient of the dynamics whose general effects on such a spontaneous symmetry breaking process are not yet fully clear and could be worth investigating.

A further interesting study not carried out in the present paper is to replace strong ideal projective measurements with weak measurements due to the interaction with a probe system. One expects that weak measurements are not effective in inducing a Zeno-like effect because the change (or ``collapse'') of the system's quantum state occurs only gradually. However, in this case, diverse probe systems can be coupled to the target system, and different probe measurements can be chosen, yielding different post-selection effects on the target system. The possibility that also a well designed weak measurements could be able to prevent phase diffusion is an interesting option which will be investigated in the future.

We finally notice that it has been proved that the standard quantum Zeno effect occurs also if measurements are replaced by strong coupling~\cite{Facchi2001,Facchi2001PRL,Xu2011}. Whether these similar phenomena can be extended to self-sustaining systems can be an interesting subject for future investigations.

\begin{acknowledgements}
We acknowledge the support of the European Union Horizon 2020 Programme for Research and Innovation through the Project No. 732894 (FET Proactive HOT) and the Project QuaSeRT funded by the QuantERA ERA-NET Cofund in Quantum Technologies. NES acknowledges the TRIL support of the Abdus Salam International Centre for Theoretical Physics (ICTP). WZZ is supported by the National Natural Science Foundation of China (Grant Nos. 12074206) and K C Wong Magna Fund in Ningbo University, China.
\end{acknowledgements}


\begin{thebibliography}{35}
\bibitem{exp2} Here the sign $\mathcal{N}(x_0,y_0)$ denotes a Gaussian distribution whose expectation is $x_0$ and variance is $y_0$, that is, the probability density function is $f(x)=(2\pi y_0)^{-1/2}\exp[-(x-x_0)^2/(2y_0)]$.
\bibitem{explain1} Note that in normal linear model one have $I_0=0$ and $\delta I_j=0$ which leads the $\dot{\theta}_-=-\Delta \omega+N_{\theta-}$ with a simple potential $U=\Delta\theta$ and no synchronization occurs.
\end{thebibliography}
\end{document}